\newif\ifanon
\anonfalse

\newif\ifepigraph
\epigraphfalse

\documentclass[letterpaper,twocolumn,10pt]{article}
\usepackage{usenix-2020-09}

\pdfoutput=1

\usepackage{amsfonts}
\usepackage{amsmath}
\usepackage{amsthm}
\usepackage{color}
\usepackage{graphicx}
\usepackage{microtype}
\usepackage{listings}
\usepackage{epigraph}
\usepackage{float}
\usepackage{caption}
\usepackage{tabularx}
\usepackage{array}
\usepackage{wrapfig}

\newcommand{\ints}{\mathbb{Z}}

\newcommand{\dotcup}{\ensuremath{\mathaccent\cdot\cup}}
\newcommand{\etal}{et al.}

\ifanon
\newcommand{\taffy}{stretchy}
\newcommand{\Taffy}{Stretchy}
\newcommand{\TBF}{SBF}
\newcommand{\TCF}{SCF}
\newcommand{\MTCF}{MSCF}
\else
\newcommand{\taffy}{taffy}
\newcommand{\Taffy}{Taffy}
\newcommand{\TBF}{TBF}
\newcommand{\TCF}{TCF}
\newcommand{\MTCF}{MTCF}
\fi

\begin{document}

\ifanon
\title{\Large \bf \Taffy{} Filters: Growable Bloom and Cuckoo Filters}
\else
\title{\Large \bf Stretching Your Data With \Taffy{} Filters}
\fi

\ifanon
\else
\author{\rm Jim Apple}
\fi





\maketitle

\begin{abstract}
Popular approximate membership query structures such as Bloom filters and cuckoo filters are widely used in databases, security, and networking.
These structures represent sets approximately, and support at least two operations -- insert and lookup; lookup always returns true on elements inserted into the structure; it also returns true with some probability $0 < \varepsilon < 1$ on elements {\em not} inserted into the structure.
These latter elements are called false positives.
Compensatory for these false positives, filters can be much smaller than hash tables that represent the same set.
However, unlike hash tables, cuckoo filters and Bloom filters must be initialized with the intended number of inserts to be performed, and cannot grow larger --
inserts beyond this number fail or significantly increase the false positive probability.
This paper presents designs and implementations of filters than can grow without inserts failing and without meaningfully increasing the false positive probability, even if the filters are created with a small initial size.
The resulting code is available \ifanon\else on GitHub \fi under a permissive open source license.
\end{abstract}



\section{Introduction}






The Bloom filter is a ubiquitous data structure that allows storing a set with a low amount of space.
Bloom filters support the operations insert -- which adds an item to the set -- and lookup, which returns true if an element is in the filter; if an element is not in the filter, true is returned with some configurable probability $0 < \varepsilon < 1$.
This is called the ``false positive probability'', or ``fpp''.

There are a number of other structures also supporting insert and lookup with a false positive probability greater than 0~\cite{vacuum,morton-journal,ribbon,xor-filter,quotient-filter,broom,vector-quotient}.
A lookup operation with these guarantees is sometimes called an ``approximate membership query'', and structures that support approximate membership queries are sometimes referred to ``AMQ structures'' or just ``filters''.
The significant interest in filters is reflective of their utility in applications such as databases, security, and networking~\cite{split-bloom, vacuum, quotient-filter, malware, profile-similarity, invertible, flooding-filter, summary-cache, prefix-matching-filter}.


Each of the filter structures cited above supports approximate membership queries on sets with a given maximum size, but the question of extensible (or {\itshape extendable} or {\itshape incremental} or {\itshape growable}) filters that can increase in capacity as more elements are inserted is little studied.
The classic answer is to create a sequence of filters, possibly of increasing sizes and/or lower false positive probabilities~\cite{dynamic-bloom,scalable-bloom,dynamic-cuckoo}.
Inserts occur on the last filter to be created and lookups must search each filter.
Even in designs for which this keeps the false positive rate low, lookup times balloon from constant to poly-logarithmic or even linear in $n$, the number of elements inserted~\cite{psw,logarithm,consistent-cuckoo}. 
Additionally, the space usage often grows as $\Omega(n \lg n)$, at which point a traditional hash table would do the same work in the same space with constant-time operations and an $n^{-c}$ false positive probability, where $c$ depends on the constant in $\Omega(n \lg n)$.
A newer approach to manage growing filters is to use cuckoo or quotient filters in which, each time the filter capacity grows, the false positive probability doubles~\cite{logarithm,morton-journal,vacuum,rsqf,entry-extensible}.
Finally, a third approach to the problem of growing a filter is to depend on the original keys being available during rebuild time~\cite{elastic}.
This approach is not always possible or time efficient.
See Figure~\ref{prior-work-table}, which describes prior work and its limitations when filters grow.

\begin{figure*}
\begin{tabular}{|m{2in}|m{4.667in}|}
\hline {\bf Behavior} & {\bf Filters} \\
\hline $\omega(1)$ lookup & dynamic bloom \cite{dynamic-bloom}, scalable bloom \cite{scalable-bloom}, dynamic cuckoo \cite{dynamic-cuckoo}, monkey \cite{monkey}\\
\hline Doubled fpp when capacity doubles & logarithmic dynamic cuckoo \cite{logarithm}, Morton \cite{morton-journal}, vacuum \cite{vacuum}, rank select quotient \cite{rsqf}, consistent cuckoo \cite{consistent-cuckoo}, dynamic cuckoo \cite{dynamic-cuckoo}, entry-extensible cuckoo \cite{entry-extensible} \\
\hline Depend on storing $\Omega(\lg n)$ bits per element (in filter or in backing store) & elastic \cite{elastic}, consistent cuckoo \cite{consistent-cuckoo}, Chucky \cite{chucky} \\
\hline More than double fpp when full beyond capacity & Bloom~\cite{bloom}, tinySet~\cite{tinyset} \\
\hline
\end{tabular}
\caption{The filter types that exhibit various undesirable behavior as more keys are inserted
\label{prior-work-table}}
\end{figure*}

Instead of these approaches, this paper investigates practical structures that allow the structure to grow and keep a low false positive rate (not exceeding a threshold specified when the structure was created), all while using no more than $O(\lg \lg n + \lg (1/\varepsilon))$ bits of space per element~\cite{psw}.
This is a significant improvement over the status quo in which filters either cannot grow, such as standard Bloom filters, or use $\lg (1 / \varepsilon) + \omega(\lg \lg n)$ or $\omega(\lg(1/\varepsilon))$ bits to represent sets with size $n$ and false positive probability $\varepsilon$.

\subsection{Applications}

Growable filters are potentially useful in situations where there is no known bound on the number of keys to be inserted.
One example is in joins in query processing systems.
It is often beneficial to performance to create and populate a filter for the build side of a join:
the filter, being much smaller than the full output of the build-side hash table construction, can be pushed-down to the probe side to reduce the number of rows that need to be tested against the build output~\cite{tpch-filter}.
If there are any predicates on the build side, or if the build side has incomplete or inaccurate distinct value count statistics, it is not possible to predict the eventual size of the filter.
Systems like Apache Impala estimate the cardinality when initializing the filter and then discard the filter if the estimate was too low~\cite{impala}.
Using growable filters would allow these filters to continue to be populated and used in the probe side.


Another example where growable filters are useful is in log-structured merge trees (``LSM trees'')~\cite{lsm}.
Log-structured merge trees store data in sorted ``runs'' of exponentially-increasing size.
In order to cheaply discover if a key is present in a run, systems like RocksDB equip each run with a filter~\cite{lsm, ribbon}.
Point lookups that go through the filters require accessing $\lg n$ filters, where $n$ is the number of keys in the LSM tree.
A single growable filter structure can reduce this to a single filter query by storing one structure for all keys, rather than $\lg n$ structures.
Here a Bloomier filter (sometimes called a retrieval data structure) is called for, in which every positive result from a lookup operation has an attached value~\cite{bloomier}.
For LSM trees, that value should be the identifier of the most-recently-created run a key is associated with.
Upon a positive lookup in the filter, the run identifier is retrieved, and a more expensive probe of that run can begin.\footnote{
The ``Chucky'' system is built on this premise, but requires a full filter rewrite at each last-level compaction~\cite{chucky}.
SlimDB also uses a cuckoo filter to implement a retrieval structure on the most-recently-created ``sub-level'' that a key is in in an LSM, but doesn't use dynamic sizing at all~\cite{slim}.
}

A final example is previously-used passwords~\cite{opus}.
The goal of a filter for these cases is to allow lookups during password creation time and prevent users from using a previously used password.
These sets can have long lifetimes and grow arbitrarily large; the ``Have I Been Pwned'' data set is 11GB of SHA-1-hashed passwords~\cite{pwned}.
Because of password databases' propensity for growth, static-capacity structures like Bloom filters or cuckoo filters are less well suited for these data sets.
Section~\ref{hibp} discusses this example in more detail.





\subsection{Contributions}

To address the need for filters that can grow, this paper makes three contributions.

\begin{enumerate}
\item Section~\ref{tbf}  presents the {\em \taffy{} block filter}          (``\TBF{}''),  a Bloom-filter-backed  AMQ structure with $O(\lg n)$ lookup cost.
\item Section~\ref{tcf}  presents the {\em \taffy{} cuckoo filter}         (``\TCF{}''),  a cuckoo-hashing-based AMQ structure with $O(1)$ lookup cost.
\item Section~\ref{mtcf} presents the {\em minimal \taffy{} cuckoo filter} (``\MTCF{}''), a cuckoo-hashing-based AMQ structure that decreases the space needed in a \TCF{} by up to a factor of 2.
\end{enumerate}

\TCF{}s, in addition to having $O(1)$ lookup, contribute a new understanding of cuckoo filters as dictionaries.
\MTCF{}s apply for the first time the technique of quotienting to dictionaries that can grow without doubling in size, which may be of independent interest.

Section~\ref{eval} describes experimental performance results on all three \taffy{} filters and what circumstances each is suited for. 
Section~\ref{conclusion} concludes.


\section{Prior work}

\begin{figure}[t!]
\begin{tabular}{|m{0.5in}|m{2.5in}|}
\hline {\bf Symbol} & {\bf Usage} \\
\hline $a$ & The logarithm, base 2, of the number of buckets in an array in a \TCF{} or an \MTCF{}. \\
\hline $b$ & The number of slots in a bucket in a filter or hash table that uses buckets. \\
\hline $B$ & The size of a block in a block Bloom filter. \\
\hline $d$ & The over-provisioning per key - the number of bits per element that need to be stored beyond $\lg (1/\varepsilon)$. \\
\hline $F$ & The size of fingerprints in \TCF{}s and the size of large fingerprints in \MTCF{}s.
See Sections~\ref{tcf}~and~\ref{mtcf}. \\
\hline $k$ & The number of hash functions in a cuckoo hash table or Bloom filter. \\
\hline $L$ & The size of a ``lane'' in a split Bloom filter. \\
\hline $m$ & The number of bits in a Bloom filter. \\
\hline $n$ & The number of keys in a filter or dictionary at a given point in time. \\
\hline $N$ & The maximum number of keys that will ever be in a filter.
Always less than $|U|$. \\
\hline $p$ & The logarithm, base 2, of the number of levels in an \MTCF{}. \\
\hline $S_i$ & The set of permutations on the integers in $[0, i)$. \\
\hline $T$ & The maximum size of tails in \TCF{}s and \MTCF{}s.
See Sections~\ref{tcf}~and~\ref{mtcf}. \\
\hline $U$ & The ``universe'' - the set of keys that could be put in a filter. \\
\hline $\ints_i$ & The set of integers $[0, i)$. \\
\hline $\ints_2^i$ & The set of bit strings of length $i$. \\
\hline $\delta$ & The over-provisioning per structure - the percent of empty space in a dictionary or filter. \\
\hline $\varepsilon$ & The false positive probability, or ``fpp''. \\
\hline $\varphi_i$ & The permutations associated with side $i$ of a \TCF{}.
See Section~\ref{tcf}. \\
\hline $A \dotcup B$ & The tagged union of $A$ and $B$ such that even if $A \subseteq B$, $A \dotcup B \ne B$. \\
\hline
\end{tabular}
\end{figure}

\subsection{Split block Bloom filters}



The insert and lookup operations in standard Bloom filters access $\lg (1/\varepsilon)$ bits in an array of size $m$ that stores $m \ln 2 / \lg(1/\varepsilon)$ distinct elements~\cite{bloom-original}.
These cause $\lg (1/\varepsilon)$ memory accesses and require the same number of hash function applications.
Block Bloom filters reduce the number of memory accesses to 1 at a cost of a slightly increased false positive probability~\cite{block-bloom}.

\begin{figure}[b!]
  \includegraphics[width=3.333in]{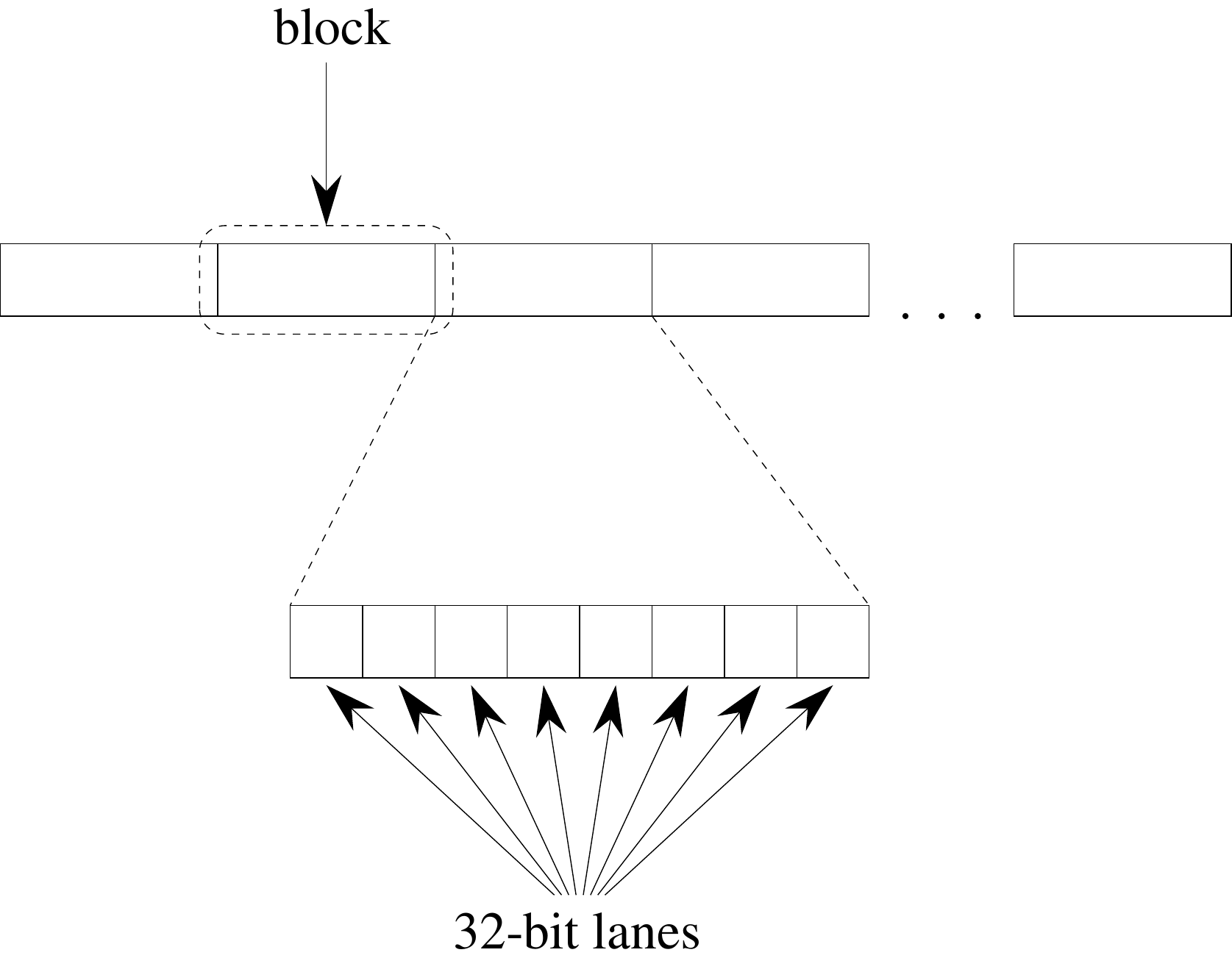}
\caption{\label{sbbf-diagram}
A diagram of a split block Bloom filter with $k = 8$ and $B = 256$.
}
\end{figure}

Each block Bloom filter is implemented as an array of non-overlapping blocks; see Figure~\ref{sbbf-diagram}.
Each block is itself a Bloom filter.
Blocks are no larger than a single cache line in size.
To insert a key, the key is hashed to select the block to use, mapping a key $x$ to $h(x) \bmod m/B$, where $h$ is the hash function, $m$ is the size of the block Bloom filter and $B$ is the size of each block.

In split block Bloom filters, once a block is selected, it is used as a ``split'' Bloom filter~\cite{split-bloom}.
In a standard Bloom filter, to insert a key $x$, $k = m \ln 2 / n$ hash functions are applied to $x$, and each bit $h_i(x) \bmod B$ is set, $0 \le i < k$.
In a split Bloom filter, the filter is split into equal-sized non-overlapping ``lanes'', each of size $L$.
Upon insertion, the bits $i L + (h_i(x) \bmod L)$ for $0 \le i < k$ are set; in other words, a single bit is set in each lane.

When a block Bloom filter is used with block size $B = 256$ and lane size $L = 32$, it is possible to use SIMD instructions to perform the eight hash function computations at once, set the eight bits at once (one per 32-bit lane), or check those eight bits at once.
The resulting Bloom filter has constant-time branch-free insert and lookup and is consistently faster than a cuckoo filter of the same size (See Figures~\ref{lookup-both}~and~\ref{arm-lookup-both} in Section~\ref{eval})~\cite{cuckoo-filter-github,ultra-fast,overtakes,impala-bloom}.

{\bf \Taffy{} block filters use split block Bloom filters as a building block to make an extensible filter with lower query time than a traditional Bloom filter would require in the same application.}

\subsection{Cuckoo hashing}

{\bf \TCF{}s and \MTCF{}s are based on cuckoo hashing, a method of collision resolution in open-address\-ing hash tables that assigns each key a small set of slots it can occupy~\cite{cuckoo-journal}.}
A cuckoo hash table consists of two arrays of size $(1 + \delta)n/2$ to store a set of $n$ keys, for $0 < \delta < 1$.
The arrays are broken up into contiguous non-overlapping buckets~\cite{buckets,load-thresholds}.
Each key is assigned one bucket per array via the application of two hash functions on the key.
Every key in the table will be stored in a slot in one of those two buckets.

Inserting a key is more complex.
If no slot in the two buckets for storing a key is empty, one of the occupying keys is evicted and replaced by the key being inserted.
Now the victim of the eviction is in turn inserted.
With high probability, eventually the evictions find an empty slot and the chain of evictions ends~\cite{cuckoo-journal}.

\subsection{Succinct dictionaries with quotienting}
\label{quotienting}

\begin{figure}[b!]
\includegraphics[width=3.333in]{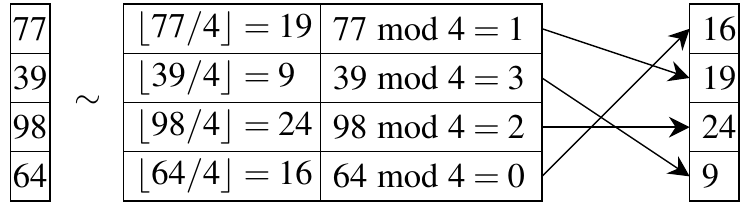}
\caption{\label{quotienting-figure}
Quotienting with $n=4$ and all buckets holding exactly one element.
}
\end{figure}

Maps of size $n$ with keys from a ``universe'' of size $U$ can be na\"ively stored in $n \lg |U|$ bits by storing every element (in any order) in an array of size $n$.\footnote{A ``universe'' is the set of all possible keys, such as all 64-bit integers, or all strings up to length 1 trillion characters.}
Space can be saved using a technique called ``quotienting''~\cite{knuth,quotient-filter}.
See Figure~\ref{quotienting-figure}.

The basic construction can be illustrated as follows:
first, an array of size $n$ is created in which each array slot can hold an arbitrary number of keys~\cite{raman-practical}.
Then, a key $x$ is stored in slot $x \bmod n$.
Additionally, instead of storing $x$ explicitly, $\lfloor x / n \rfloor$ is stored;
$x \bmod n$ is the {\em implicitly-stored} part of they key and $\lfloor x / n \rfloor$ is the {\em explicitly-stored} part of the key.
Because only $\lfloor x / n \rfloor$ is stored as the key, only $\lg |U| - \lfloor \lg n \rfloor$ bits are required to store it.
Coming back to the array, this reduces the total storage required to $n (\lg |U| - \lfloor \lg n \rfloor)$.

In Figure~\ref{quotienting-figure}, the column on the left represents a set of values in $\ints_{128}$ (integers between 0 and 127, inclusive), with each element taking 7 bits to store.
The column in the middle shows another way of representing the same set as two parts per element: one of the lower order two bits and another of the higher order five.
The column on the right stores the two low-order bits implicitly and the high order five bits explicitly.
This cuts the space needed to store the set down from 28 bits to 20 bits.

{\bf \TCF{}s and \MTCF{}s use quotienting in cuckoo hashing to reduce the space needed to store the filter.}

\subsection{Filters that can grow}
\label{filters-that-grow}

Pagh \etal{} describe two constructions to support extensible filters~\cite{psw}.
{\bf \Taffy{} filters refine the work of Pagh \etal{} with new structures for both constructions.}

\paragraph{$O(\lg n)$ lookup}
The first is implemented as a series of succinct dictionaries.
Common similar constructions use a series of Bloom filters and exponentially decreasing false positive probabilities in each subsequent filter in order to bound the total false positive rate.
That is, they create a sequence of Bloom filters with the following pairs for the false positive probability and expected number of distinct values:

\[
\langle \varepsilon / 2, 2 \rangle,
 \langle \varepsilon / 4, 4 \rangle,
 \langle \varepsilon / 8, 8 \rangle,
 \langle \varepsilon / 16, 16 \rangle,
 \ldots
\]

As new items arrive, they are inserted into the largest Bloom filter.
Once that filter reaches the capacity it was configured for, a new Bloom filter with twice the capacity and half the false positive probability is initialized.
Lookups access all the Bloom filters.

\begin{figure}[b!]
\includegraphics[width=3.333in]{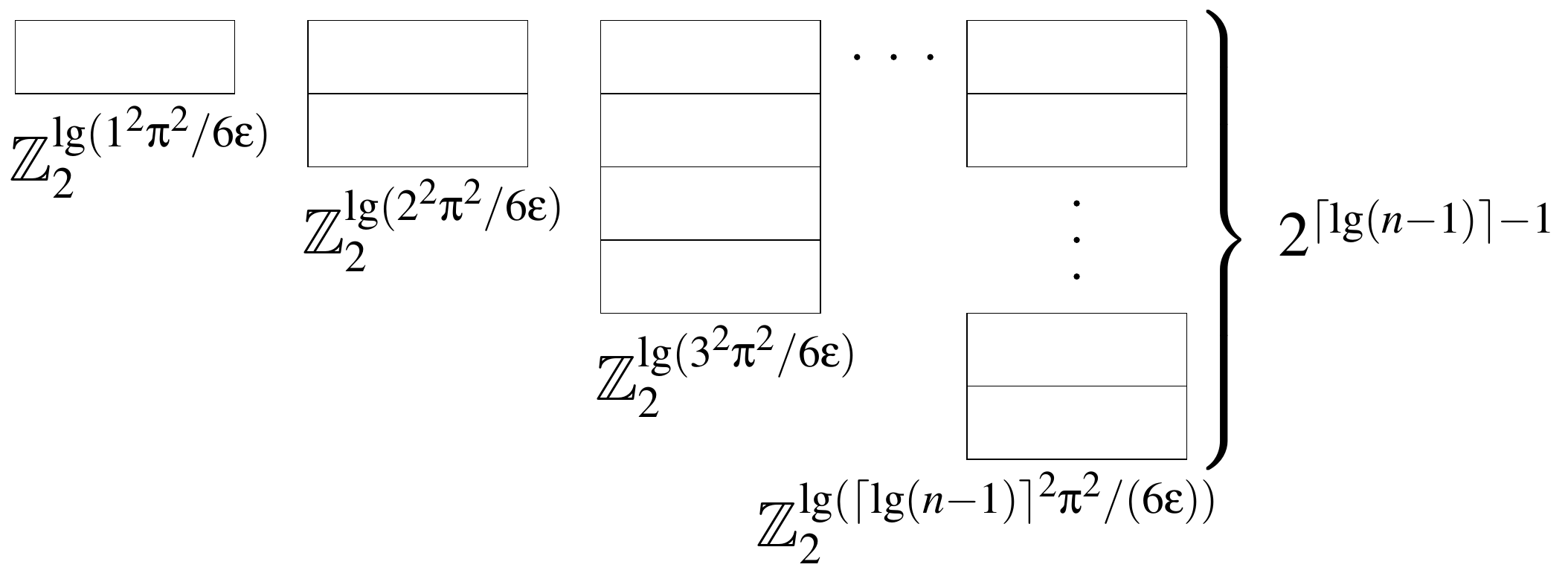}
\caption{\label{pagh-1-diagram}
Pagh \etal{}'s first construction.
$\ints_2^m$ means bitstrings of length $m$.
In this diagram, columns of blocks represent dictionaries.
The caption under a column is the type of the elements in the dictionary. For instance, the block with four rows in its column stores bit strings of length $\lg (3^2 \pi^2 / 6 \varepsilon)$.
Quotienting (see Section~\ref{quotienting}) and other space factors are not presented in this figure.
}
\end{figure}

This leads to a storage footprint of more than $(\lg n + \lg (1/\epsilon)) / \ln 2$ bits per element and a query time of $O(\lg^2 n + \lg n \lg (1/\varepsilon))$.
Pagh \etal{} reduce the lookup cost to $O(\lg n)$ by using a dictionary like Raman and Rao's that has $O(1)$ query time per filter~\cite{psw,succinct}.
They also reduce the space usage to $O(\lg \lg n + \lg (1/\varepsilon))$ bits per element by using the sequence $\langle O(\varepsilon / i^2),  2^i \rangle$ for $i \in [1,\infty)$, rather than  $\langle \varepsilon / 2^i,  2^i \rangle$.
See Figure~\ref{pagh-1-diagram}.
In this construction, $\lceil\lg (n-1) \rceil$ dictionaries are maintained with exponentially increasing capacities and logarithmically increasing bit widths.
The false positive probability of the $i$th dictionary, counting from $1$, is $6 \varepsilon / i^2 \pi^2$, and the sum of the false positive probabilities is $\le \varepsilon$.
The lookup operation requires a dictionary lookup in $\lceil\lg(n-1)\rceil$ dictionaries.

\begin{figure}[b!]
\includegraphics[width=3.333in]{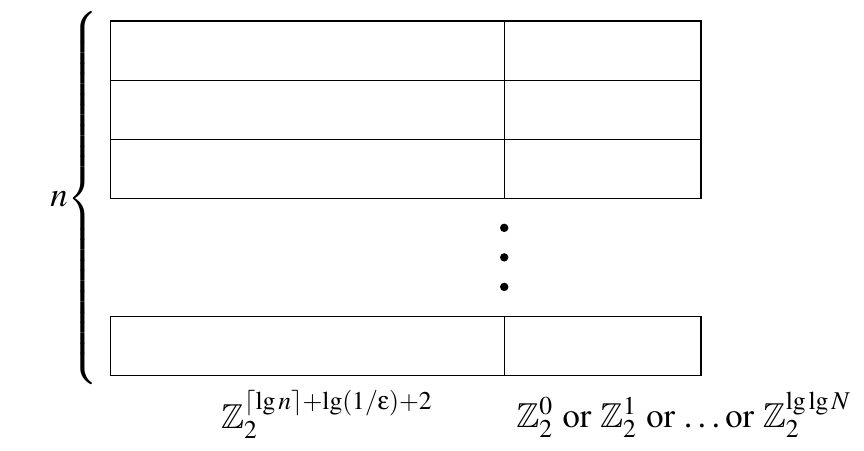}
\caption{\label{pagh-diagram}
Pagh \etal{}'s second construction.
In this construction of a growable filter, when a filter contains $n$ items, it is stored as a dictionary in which the keys are bit stings of length $\lceil \lg n \rceil + \lg (1/\varepsilon) + 2$ and the values are bit strings of length up to $\lg \lg N$, where $N$ is the largest number of keys the filter will contain.
}
\end{figure}

\paragraph{$O(1)$ lookup}
Pagh \etal{} also present a filter with the same space usage but $O(1)$ query time~\cite{psw}.
See Figure~\ref{pagh-diagram}.

This filter maintains a map where the keys are bit strings of length $\lceil \lg n \rceil + \lg (1/\varepsilon) + 2$ and the values (which we will call ``tails'') are bit strings of length up to $\lg \lg N$, where $N$ will be the largest size of the data structure.
(This definition of $N$ is not a problem in practice, as using $|U|$, the size of the universe of keys, should be sufficient for integer keys. For non-integer keys, they must be hashed down to an integer in order to use these structures, and using the universe of the set of integers each key is hashed to also works well.)
After every $2^i$ insertions, a new map is created where the keys are one bit longer.
Pagh \etal{} show that the fpp of such a dictionary is no more than $\varepsilon$ as long as $n < N$.

\subsection{Compact extensible dictionaries}

Hash tables that are used to accommodate sets without a size known in advance typically do so by doubling in capacity as needed.
This applies to \TCF{}s, as well.
This means that at least 50\% of the space goes unused at points, with an average unused percentage of at least 25\%.
Constructions like that of Raman and Rao are able to mitigate this, but they are largely theoretical~\cite{succinct}.
Instead, Maier \etal{} use the cuckoo hashing evict operation to incrementally resize a hash table~\cite{dysect}.
First, the ``DySECT'' table, as they call it, is broken up into equal sized sub-arrays that can be resized independently.
When the table gets close to full, exactly one of the sub-arrays is doubled in size.
This frees up room that's available in future eviction sequences, and the new space will slowly be filled.
Eventually all arrays will have been doubled in size, thereby causing the whole table to have doubled in size without going through a phase with as low as 50\% space usage.

{\bf \MTCF{}s filters extend quotienting-based dictionaries to DySECT tables for the first time.}

\section{\Taffy{} block filters}
\label{tbf}

The first construction from Pagh \etal{} consists of a set of sub-filters of geometrically decreasing false positive probabilities but exponentially increasing size~\cite{psw}; see Figure~\ref{pagh-1-diagram}.
As Pagh \etal{} describe it, this filter is initialized with a single sub-filter.
Inserts take place on the most recently added sub-filter (which is the largest), while lookups are performed by performing a lookup in each sub-filter until the element is found or there are no more sub-filters to search.
Once $2^i$ inserts have taken place, a new sub-filter is initialized and added to the collection.

Using traditional Bloom filters, the lookup cost would be

\[
\begin{array}{r c l}
\displaystyle\sum_{i=1}^{\lg n} \lg (i^2 \pi^2 /(6 \varepsilon)) & = &
 \displaystyle\sum_{i=1}^{\lg n} \lg (i^2) + \lg( \pi^2) - \lg 6 + \lg (1/\varepsilon) \\
& = & \Theta(\lg^2 n + \lg n \lg (1/\varepsilon))
\end{array}
\]

Instead, Pagh \etal{} use dictionary-based filters that support constant-time lookup -- such as Raman and Rao's dictionary -- rather than Bloom filters~\cite{succinct,psw}.
This reduces the lookup time to $O(\lg n)$.

{\bf \Taffy{} block filters} (``\TBF{}s'') use split block Bloom filters to keep the lookup time logarithmic and independent of $\varepsilon$, rather than Raman and Rao's dictionary, as the latter is a theoretical, not a practical design~\cite{psw,succinct}.
Split block Bloom filters have proven to be the fastest dynamic filters for doing single-element lookups in recent works on the matter~\cite{overtakes,ribbon,bloom-simd}.
Even though they are not based on dictionaries, they suffice for this construction, as they support the two operations needed for each level: lookup and insert.
See Section~\ref{eval} for performance of \TBF{}s compared to pre-sized split block Bloom filters.


\section{\Taffy{} cuckoo filters}
\label{tcf}

\Taffy{} block filters' lookup operation requires $\lg n$ lookup operations on their sub-filters, one per sub-filter.
\Taffy{} cuckoo filters reduce lookup times to $O(1)$ and show how the ideas from quotienting can be applied to cuckoo filters to produce a dictionary.

\Taffy{} cuckoo filters (``\TCF{}s'') use quotienting cuckoo tables 
to store their data, as this reduces the storage space by a significant margin (See section~\ref{quotienting}).
See Figure~\ref{pagh-diagram}.
The keys are bit-strings of length $\lfloor \lg n \rfloor + F$ and the values are bit-strings of length up to $T$, for some fixed $F$ (for ``fingerprint'') and $T$ (for ``tail'').
By quotienting in an array of size $\Omega(n)$, each fingerprint-tail pair can be stored in $\lfloor \lg n \rfloor + F + T - \lg n + O(1)$ bits, for a total space usage of $(F+T)n + O(n)$.
For performance and simplicity purposes, we pick $F + T = 15$, but this is not a requirement of the structure.




\paragraph{Quotienting} Quotienting is used with linear probing as the collision resolution mechanism in quotient filters~\cite{quotient-filter}.
Quotienting can also be used with cuckooing as the collision resolution mechanism, as in backyard cuckoo hashing~\cite{backyard}.
Cuckoo hash tables maintain $k \ge 1$ potential locations for each key, each of which could be stored in any of its potential locations~\cite{cuckoo-journal}.
Because more than one hash function is used and because eviction occurs, it must be possible to translate from a location-element pair to an alternate location-element pair for the same key.
See Section~\ref{dictionary}.





\TCF{}s are based on cuckoo tables and have two arrays of slots, referred to as ``sides,'' just as (some) cuckoo filter designs break up the address space into multiple regions, one per hash function.
With \TCF{}s, this is a requirement in order to be able to recover enough of the original key in order to re-hash it to a larger address space.
\TCF{}s, unlike backyard cuckoo hashing, use bucketing in order to increase the usable capacity and thus reduce wasted space~\cite{backyard,buckets}.\footnote{Like quotienting, buckets are not required for the correctness of the structure, just its succinctness.}
Each side of a \TCF{} comes equipped with a random permutation on bit-strings of length $\lg n + F - O(1)$, analogous to how each side of a cuckoo hash table comes equipped with a hash function.
A fingerprint-tail pair $(f, t)$ is stored in one of two buckets: the one in side 0 pointed to by the high-order $\lg n - O(1)$ bits of $\varphi_0(f)$ or the one in side 1 pointed to by the high-order $\lg n - O(1)$ bits of $\varphi_1(f)$, where $\varphi_i$ is the permutation associated with side $i$.

The critical part of the permutations is the ability to translate between $S_i(x)$ and $S_j(x)$ for a key $x$ and $i \ne j$.
In a cuckoo hash table in which the original key $x$ is stored, this is trivial whether or not the hash functions associated with each side are permutations.
In \taffy{} cuckoo filters this translation is accomplished without storing $x$ directly, but just $S_0(x)$ and $S_1(x)$, via $S_i(x) = S_i(S_j^{-1}(S_j(x)))$.

\paragraph{}
More concretely, see Listing~\ref{tcf-types}.
An {\em element} consists of two groups of bits.
The fingerprint (of size $F$) is tested for equality when executing the lookup operation; the tail (of size 0,1,\dots, or $T$) is the unused part of the hashed key that will eventually be used in the fingerprint (after permuting -- see below).
A {\em bucket} consists of $b$ possibly empty slots, each of which can hold one element or be empty.\footnote{
For our implementation, we use $b = 4$.
Just as with $F$ and $T$, four is not a magic number, but one picked for a balance between maximum load and fpp, both of which go up as buckets get larger.
}
A {\em side} consists of $2^a$ buckets for some $a$ as well as a random permutation on $\ints_2^{a+F}$. 
A \TCF{} consists of two sides and one hash function that produces a 64-bit key.
The two sides have the same number of buckets but different permutations.

In Listing~\ref{tcf-types},       $A \dotcup B$ represents a tagged disjoint union of $A$ and $B$; even if $A \subseteq B$, $A \dotcup B \ne B$.
      \texttt{T}$_\bot$ means the type \texttt{T} extended with the element $\bot$, indicating ``null'' or ``empty''.
      \texttt{T[$n$]} denotes an array of $n$ values of type $T$.
      $S_i$ is the symmetric group on $\ints_i$ -- the set of all permutations on $\ints_i$.
      Structs are denoted by curly brackets \{\}. 

\ifanon
\begin{lstlisting}[escapeinside={`}{`},float,label=tcf-types,
    caption={The types of a \TCF{}.
    }
    ]
Element := {fingerprint: `$\ints_2^F$`, tail: `$\dotcup_{i \le T} \ints_2^i$`}
Slot := Element`$_\bot$`
Bucket := Slot[`$b$`]
Side(a) := {Bucket[2`$^a$`], Permutation: `$S_{2^{k+F}}$`}
SCF(U, a) := {Side(a)[2], HashFunction: `$U \to \ints_2^{64}$`}
\end{lstlisting}
\else
\begin{lstlisting}[escapeinside={`}{`},float,label=tcf-types,
    caption={The types of a \TCF{}.
    }
    ]
Element := {fingerprint: `$\ints_2^F$`, tail: `$\dotcup_{i \le T} \ints_2^i$`}
Slot := Element`$_\bot$`
Bucket := Slot[`$b$`]
Side(a) := {Bucket[2`$^a$`], Permutation: `$S_{2^{k+F}}$`}
TCF(U, a) := {Side(a)[2], HashFunction: `$U \to \ints_2^{64}$`}
\end{lstlisting}
\fi

A slot is encoded in a bitfield of size $F+T+1$ as follows.
If the last $T+1$ bits are all zero, the slot is empty.
Otherwise, there must be a one bit in the last $T+1$ bits.
All bits following that one bit are the tail.
Bits $1-F$ are the $F$-bit fingerprint.
For example, the tail {\tt 010000} represents a tail of size 4, {\tt 0000}, while {\tt 000001} represents tail of length zero.
The tail {\tt 000000} represents an empty slot, which is dinstinct from an element with a tail of length zero.

\paragraph{Lookup}
A lookup begins by hashing a key with the \TCF{}'s hash function. 
Then the lookup operation does the following:

\begin{enumerate}
\item Applies the permutation associated with side 0, $\varphi_0$, to the most-significant $a + F$ bits in the key. 
\item Reserves the next $T$ bits of the key; this will be the key's tail.
Note that these bits have not been permuted.
\item Using the most-significant $a$ bits in the permuted bits, selects a bucket within side 0. 
(The remaining $F$ bits in the permuted value are the fingerprint.)
\item Checks to see if one of the $b$ slots in the bucket contains an identical fingerprint. 
If so, checks if the element's tail is a prefix of the key's tail.
If yes, returns \verb|True|.
Otherwise, repeats with side 1.
If neither side contains an identical fingerprint and prefix-matched tail, returns \verb|False|.
\end{enumerate}

Note that the prefix check is not strictly necessary, but does serve to reduce the fpp.
See Upsize, below, as well as Figure~\ref{ideal-bits-per-item} in Section~\ref{eval}.

\begin{figure}[b!]
  \includegraphics[width=3.333in]{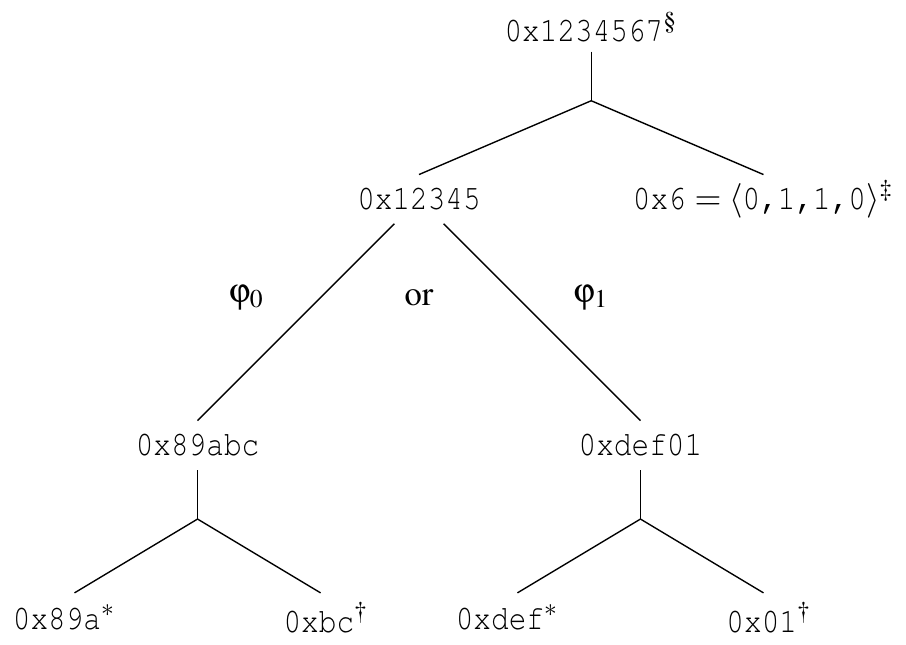}
\caption{\label{tcf-key-split}
\TCF{} key split.
In this example, $a = 12$, $F=8$, $T=4$, $\varphi_0(${\tt 0x12345}$) = ${\tt 0x89abc}, and $\varphi_1(${\tt 0x12345}$) = ${\tt 0xdef01}.\\
${}^\S$ hashed key\\
${}^\ddag$ tail \\
${}^\dag$ fingerprint \\
${}^*$ index into bucket array; stored implicitly using quotienting
}
\end{figure}

Figure~\ref{tcf-key-split} shows how a hashed key is broken down into three parts: the index into the bucket array, the fingerprint, and the tail.

\paragraph{Insert}
Insert places the key's fingerprint and tail in one of the $2b$ slots corresponding to that key, if an empty slot is found.
Otherwise, insert selects an occupied slot from the bucket to {\em evict}: the element in this slot will be moved to the other side. 

The evict operation first reconstructs the high order $a + F$ bits of the key by concatenating the $a$ bits of the bucket index and the $F$ bits of the fingerprint, then applying that side's permutation in reverse to the value. 
Using the same tail (this does not get permuted), the evict operation then inserts the evicted data into the opposite side;
this continues until an empty slot is encountered 





\paragraph{Upsize}
When a \TCF{} is nearly full, inserts may fail.
This is identical to the situation with cuckoo filters.
When this happens, the upsize method must be called to double the size of the structure. 
(This is not available in cuckoo filters without shortening the fingerprints, thereby doubling the fpp.)

The upsize operation begins by creating a new \TCF{}. 
To transfer the data from the older to the newer \TCF{}, upsize uses a modified version of the evict algorithm, as follows:

Upsize first reconstructs the $a+F$ bits of the key that were used to construct the bucket index and fingerprint. 
Then a bit is ``stolen'' from the tail and appended onto the end of the key. 
The high order bit of the tail is removed from the tail and added to the low-order end of the key.
Since the tail was taken unaltered from the key, this gives $a+F+1$ bits of the original key.
The new tail has now been decreased in length by one.
The key and this new tail can now be inserted into one of the sides of the new \TCF{} as described above. 

This works as long as the tail has positive length.
If the tail has length zero, there is nothing to steal from.
Instead, two candidate keys are created from the reverse-permuted $a+F$ bits by appending a zero and a one. 
It's indeterminate which one of these was in the original key, so both are inserted. 

The fpp remains less than $2^{-F+O(1)}$:
after adding $n$ elements to the filter, the filter holds $n/2$ fingerprint-tail pairs with tail length $\lg \lg N$, $n/4$ pairs with tail length $\lg \lg N - 1$, \dots and $n/\lg N$ pairs with tail length 1.
It also contains $(\lg n - \lg \lg N)  n / 2 \lg N$ pairs with tail length 0.
Overall that's $3n/2 - O(n)$ pairs, so the space usage is still linear in the number of elements.
(Note that if the tails started with length 0 instead of $\lg \lg N$, this would work out to $\Theta(n \lg n)$ rather than $3n/2 - O(n)$.)
Now the odds that any bitstring of length $L$ matches any of $m$ different $L$-bit strings is $m 2^{-L}$.
Applying this to \TCF{}s, since the space usage is linear, and since the ``sides'' (the two arrays of buckets) are of length $\lg n - O(1)$, then by reversing the quotienting operation we get that the probability that any random value that {\em wasn't} inserted matches with any of the existing elements is $O(n) 2^{-\lg n + O(1) - F} = 2^{-F+O(1)}$; {\em this is the false positive probability}.

Note that if the tails all started with length 0, rather than $\lg \lg N$, then the space usage would be $\Theta(n \lg n)$ and the fpp would be $2^{-F+O(1) + \lg \lg n}$.
See also~\cite{psw}.

\paragraph{Freeze and Thaw}
\TCF{}s also support {\em freeze} and {\em thaw} operations.
Freeze reduces the space consumption of a \TCF{} from $O(\lg(1/\varepsilon) + \lg \lg N)$ to $O(\lg (1 / \varepsilon))$ bits per item, where $N$ is the largest size the structure will grow to.
It does so by recreating the structure as a \TCF{} with tail length capacity $0$.
Thaw simply turns a frozen structure into an unfrozen structure by recreating a \TCF{} with tail length capacity $\lg \lg N$ in which all of the tails have length zero.
This allows new inserts to take place while capturing their tails.

\subsection{Cuckoo filters $\cong$ cuckoo hashing with permutations and quotienting}
\label{dictionary}

Note that the frozen \taffy{} cuckoo filter is a variant of a cuckoo filter in which the fingerprint hash function takes into account the index as well.
In the original cuckoo filters, the two buckets a fingerprint could reside in are separated by a hashed value of the fingerprint~\cite{cuckoo}.
The fingerprint stored in either bucket is identical, and there is no recovery of the original hashed value.
The difference between a frozen \TCF{} and the original cuckoo filter is that a frozen \TCF{} can recover a prefix of the hashed key by way of inverting the relevant permutation and applying it to the bucket index and fingerprint.
Other than this difference, the structures have the same operations.

This isomer of cuckoo filters shows how to support in cuckoo filters a straightforward method for porting techniques that were designed for cuckoo hash tables, including satellite data (making a cuckoo filter a type of Bloomier filter), overlapping blocks, stashes, $L > 2$ buckets, fast insertion algorithms, and cuckoo hashing with pages~\cite{cuckoo-simple,cuckoo-overlap,stash,d-ary,d-ary-filter,vertical,bloomier,cuckoo-linear-insertion,cuckoo-simd-insert,cuckoo-pages,cuckoo-pages-non-contiguous}.

\section{Minimal \taffy{} cuckoo filters}
\label{mtcf}

\Taffy{} cuckoo filters suffer from a step-function space usage:
at each point, the structure has a size which is a power of two, sometimes allocating twice as much space as is needed. (See Figure~\ref{space-steps} in Section~\ref{eval}.)
Even if the size were not limited to being a power of two, as in vacuum filters or Morton filter, doubling the capacity during upsize would reduce the space utilization to less than 50\%~\cite{vacuum, morton-journal}.
To address this, this section describes a cuckooing structure based on DySECT to reduce the space usage closer to only what is needed~\cite{dysect}.

DySECT is a variant of cuckoo hashing.
A DySECT table consists of some number of subtables, and as the table gets more and more full, it grows by doubling the size of one of its subtables.
Just as in cuckoo hashing, upon an insertion, an element may be evicted.
As new elements are inserted into the table, they evict older elements, and this movement causes the newly-doubled subtable to fill up.

This section proposes minimal \taffy{} cuckoo filters (``\MTCF{}s''), an application of the DySECT idea to quotienting and \taffy{} filters.
Some complications arise:

\begin{enumerate}
  \item Because subtables have different sizes, the bits that are implicitly stored using quotienting vary depending on which part of the table an element is in.
    To address this, fingerprints in \MTCF{}s have variable size.
  \item Because fingerprints have variable size, there must be multiple permutations per side, one for each size of fingerprint.
  \item Because there are multiple permutations per side, a key may be in multiple distinct buckets per side, which decreases the lookup performance and increases the false positive probability.
\end{enumerate}

\begin{figure}[b!]
  \includegraphics[width=3.333in]{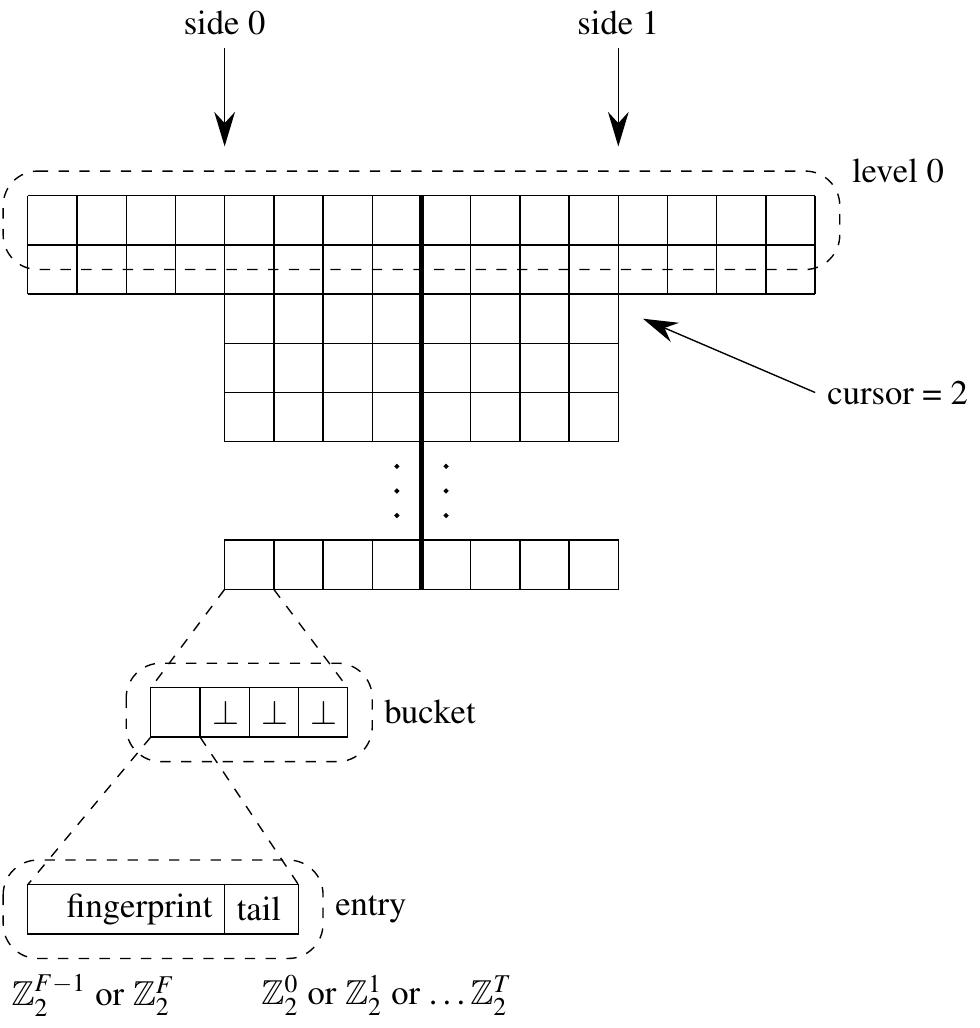}
\caption{\label{mtcf-diagram}
A diagram of an \MTCF{}.
In this example, $a = 2$ and $b=4$.
}
\end{figure}

See Listing~\ref{mtcf-types} and Figure~\ref{mtcf-diagram} for a breakdown of the components of an \MTCF{}.
In an \MTCF{}, each element has a fingerprint of size $F-1$ or $F$ and a tail of size up to $T$.
A bucket consists of $b$ (possibly empty) slots, each of which can hold one element.
A level consists of two arrays of the same size, each with $2^a$ buckets for some $a$.
The table consists of four permutations, one hash function, $2^p$ levels, and one cursor pointing to some index in the set of levels.
The maximum and minimum $a$ across all levels differ by at most 1.
Levels at location less than the cursor have the larger size.
If all levels have the same size, the cursor must be 0.

The permutations are grouped by side, two for each.
The permutations are on values with length $p + a + F - 1$ and $p + a + F$, where $2^a$ is the size of the smallest table, measured in buckets.

\ifanon
\begin{lstlisting}[escapeinside={`}{`},float,label=mtcf-types,
    caption={
      The types of an \MTCF{}.
      %A Permutation(k) is a tagged union of permutations; wlog it can be applied to bit strings of known length.
  }]
Element := {fingerprint: `$\ints_2^{F-1} \dotcup \ints_2^F$`, tail: `$\dotcup_{i \le T} \ints_2^i$`}
Slot := Element`$_\bot$`
Bucket := Slot[`$b$`]
Level(a) := Bucket[2][2`$^a$`] `$\dotcup$` Bucket[2][2`$^{a+1}$`]
Permutation(a) := `$S_{2^{p + a + F - 1}} \dotcup S_{2^{p + a + F}}$`
MSCF(U, a) := {cursor: `$\ints_{2^p}$`,
               Level(a)[2`$^p$`],
               Permutation(a)[2],
               HashFunction: `$U \to \ints_2^{64}$`}
\end{lstlisting}
\else
\begin{lstlisting}[escapeinside={`}{`},float,label=mtcf-types,
    caption={
      The types of an \MTCF{}.
      %A Permutation(k) is a tagged union of permutations; wlog it can be applied to bit strings of known length.
  }]
Element := {fingerprint: `$\ints_2^{F-1} \dotcup \ints_2^F$`, tail: `$\dotcup_{i \le T} \ints_2^i$`}
Slot := Element`$_\bot$`
Bucket := Slot[`$b$`]
Level(a) := Bucket[2][2`$^a$`] `$\dotcup$` Bucket[2][2`$^{a+1}$`]
Permutation(a) := `$S_{2^{p + a + F - 1}} \dotcup S_{2^{p + a + F}}$`
MTCF(U, a) := {cursor: `$\ints_{2^p}$`,
               Level(a)[2`$^p$`],
               Permutation(a)[2],
               HashFunction: `$U \to \ints_2^{64}$`}
\end{lstlisting}
\fi

If there are larger and smaller levels, then every element in the larger levels has a fingerprint of size $F-1$, not $F$.
This is because the implicitly-stored part of the key is one-bit longer in the larger levels, so the explicitly stored part is shorter.

In an \MTCF{}, upsize only increases the size of one of the levels, not the whole structure.
As a result, the capacity of the filter tracks more closely the number of entries in the table. (See Figure~\ref{space-steps} in Section~\ref{eval}.)

\paragraph{Lookup}
A lookup operation in an \MTCF{} first applies each of the four permutations to the hashed key.
\begin{itemize}
\item For the permutations on $p + a + F$ bits 
the first $p$ bits indicate the level, 
the next $a$ or $a+1$ indicate the bucket, 
and the remaining $F-1$ or $F$ bits are the fingerprint.
Lookup proceeds as it does in the \TCF{} case, by checking if fingerprints match and if the stored tail is a prefix of the tail of the key being looked up. 
\item For the permutations on $p + a + F - 1$ bits, the first $p$ bits again indicate the level. 
\begin{itemize}
\item If the level has tables with $2^{a+1}$ buckets, the permuted key is not used for lookup; to do otherwise would leave only $p + a + F - 1 - p - (a+1) = F-2$ bits for the fingerprint, which is not permitted.
That key is simply skipped and the lookup continues with the next key. 
\item Otherwise, the level has tables with $2^a$ buckets, and we can proceed as in the $p+a+F$ case. 
\end{itemize}
\end{itemize}




\paragraph{Insert}
In insert operations, as in lookup, the first $p$ bits of the permuted item indicate the level. 
Just as in \TCF{}s, the insert operation on a bucket may produce an eviction. 
During an evict operation in an insert, an element may move between levels with differently-sized arrays of buckets.
When the fingerprint has size $F-1$ and the level moved {\em from} has a bucket array of size $2^a$ and the level moved {\em to} has a bucket array of size $2^{a+1}$, the number of explicitly stored bits (the fingerprint bits) is now $(a + F - 1) - (a+1) = F - 2$.
Since every fingerprint must be of length $F$ or $F-1$, a bit must be stolen from the tail. 
As in \TCF{}s, if there are no bits to steal, two new key prefixes are created and inserted, as one of them must be the prefix of the original key. 

Note that \TCF{}s only steal bits during upsize operations, unlike \MTCF{}s.

\begin{figure}[b!]
  \includegraphics[width=3.333in]{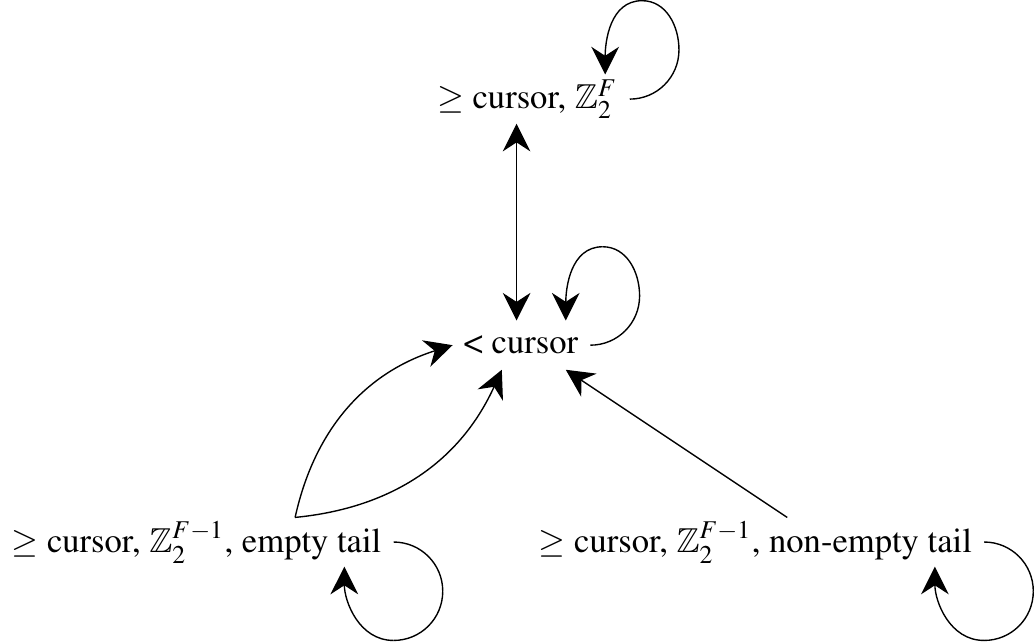}
\caption{\label{mtcf-state-transition}
The transitions an element in an \MTCF{} can go through when evicted.
}
\end{figure}

See Figure~\ref{mtcf-state-transition}, which illustrates the transitions an element in an \MTCF{} can go through when evicted.
The states indicate the lengths of the level, fingerprint, and tail.
For instance, when a level's index is less than the cursor (the center state in the diagram), the length of that level is twice what it would be if its index were higher.
For elements in a short level with a short fingerprint and a tail of length zero, when they are evicted to an element in a long level, two elements are created, as it is impossible to steal a bit from the tail of length zero.

\paragraph{Freeze/thaw and encoding}
The analysis of freeze/thaw is identical to that of \TCF{}s, with the exception that a frozen \MTCF{} is not a cuckoo filter.
It is, instead, a new type of filter that mixes cuckoo hashing and DySECT.

The tails are encoded as they are in \TCF{}s - by removing the leading zero bits and the first one bit.
An all-zero tail means a slot is unoccupied.
The fingerprint is encoded by using a single bit to indicate if the fingerprint has size $F$ or $F-1$.
Thus, a slot needs to store $F + 1 + T + 1$ bits.
For speed of operation, we choose $F = 9$ and $T = 5$ so that slots fit into {\tt uint16\_t}s.
As with \TCF{}s, and for the same reason, we pick $b$, the number of slots in a bucket, to be 4.
We pick $p$, the logarithm of the number of levels, to be $5$; other choices are valid, but this one made a nice compromise between insert speed and space overhead.



\section{Evaluation}
\label{eval}


\TBF{}s, \TCF{}s, and \MTCF{}s have been implemented and tested for correctness; this section describes their space usage, false positive probabilities, and performance.

In each chart, \TBF{}s are configured for a maximum fpp of 0.4\%.
All \taffy{} filters are configured with an initial capacity of 1. 
All experiments were performed on both an Intel i7-7800X with 96GB of memory and SMT turned on and an AWS EC2 instance of class m6g.medium with 4GB of memory and a single Graviton2 ARM-based core.
The experiments used Ubuntu 18.04 and 20.04, respectively, and g++ 10 and 9, respectively.

For performance testing, we equipped both \TCF{}s and \MTCF{}s with stashes, extra storage slots not associated with any bucket~\cite{stash}.
We set both filters to upsize when they were 90\% full or their stashes had size greater than 4.
For the random permutations we use Feistel networks with 2-independent multiply-shift as the round function~\cite{two-independent-multiply-shift}.
These are not perfectly random, of course; analyzing the sufficiency in theory is future work.~\cite{why-simple,backyard}.

For comparison, the graphs also include a cuckoo filter (labeled ``CF'') with fingerprints of size 12 and a split block Bloom filter (labeled ``SBBF'') sized to hold 100 million elements with an fpp of 0.4\%.
The keys used in the experiments are all randomly-generated 64-bit integers.
This suffices for testing larger universes as well, as noted in Section~\ref{filters-that-grow}.
To benchmark the insert time, 100 million elements are inserted.
At intermediate points we also benchmark the lookup operation one million times both on integer keys that are guaranteed to be present and on randomly-generated integer keys.
Figures~\ref{lookup-both}~and~\ref{arm-lookup-both}, which show the results of lookup performance testing, only show the randomly-generated-keys result, as the same chart for guaranteed-to-be-present keys shows the same characteristics.

\subsection{Space}

A filter with a false positive probability of $\varepsilon$ must take up at least $\lg (1/\varepsilon)$ bits per element (assuming all data sets of the same size are equally likely)~\cite{lower-bound}.
Practical filters use more space.
For instance, Bloom filters use $\lg (1/\varepsilon)/\ln 2$ bits per element, which is about $1.44 \lg (1/\varepsilon)$.
Cuckoo filters and quotient filters use $(\lg (1/\varepsilon) + d) (1 + \delta)$ where $d$ is between 2 and 3, and $\delta$ is the over-provisioning factor, between 1\% - 20\%~\cite{cuckoo,quotient-filter,vector-quotient}.
Static filters that only support a single initializing bulk insert -- such as the ribbon filter -- can use nearly optimal space~\cite{ribbon}.

\begin{figure}[b!]
  \includegraphics[width=3.333in]{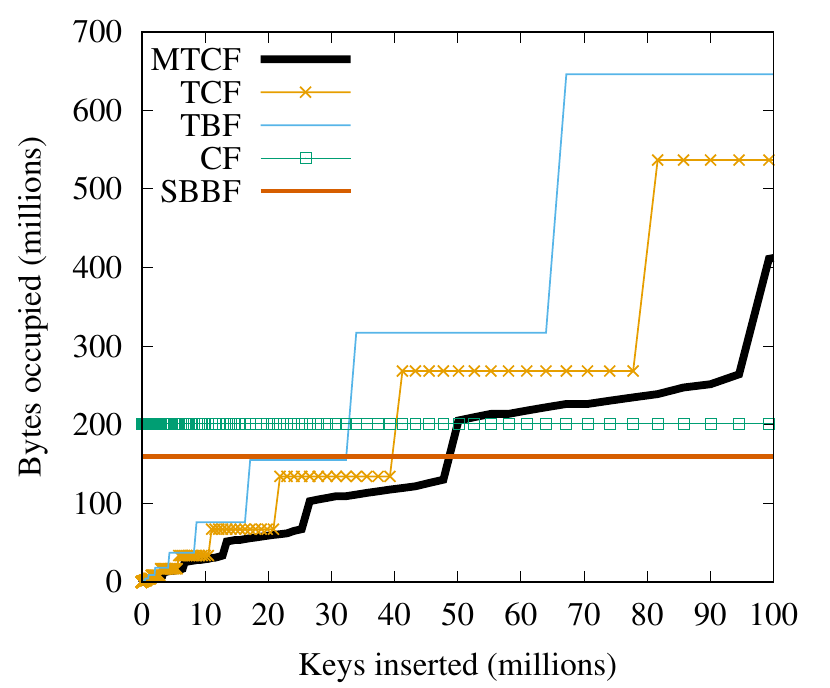}
  \caption{
    \label{space-steps}
    The amount of space used by each filter at the given number of keys inserted.
  }
\end{figure}

\begin{figure}[b!]
  \includegraphics[width=3.333in]{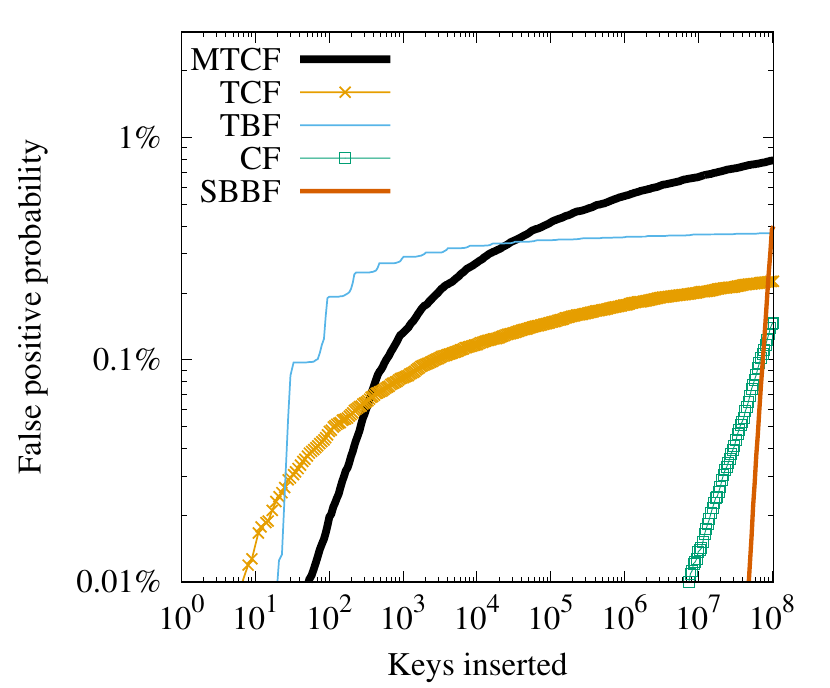}
  \caption{  \label{ideal-bits-per-item}
    $\varepsilon$, the false positive probability.
  }
\end{figure}

However, Pagh \etal{} showed that filters that can grow, like \taffy{} filters can, must use at least $\lg (1/\varepsilon) + \Omega(\lg \lg n)$ bits per element~\cite{psw}.
Figures~\ref{space-steps}~and~\ref{ideal-bits-per-item} show the space usage and $\varepsilon$, respectively.
Cuckoo filters cannot grow (without changing the number of bits per slot and doubling the false positive rate), and as such, cuckoo filters with sufficient capacity to insert up to 100 million keys use tens of millions of bytes even when the set currently stored is very small.
The same is true of split block Bloom filters.
Even though the fpp of \taffy{} filters seems to grow as the capacity grows, it is bounded above by $2^{-F+O(1)}$; see Section~\ref{tcf}.







\subsection{Time}

\begin{figure}[b!]
  \includegraphics[width=3.333in]{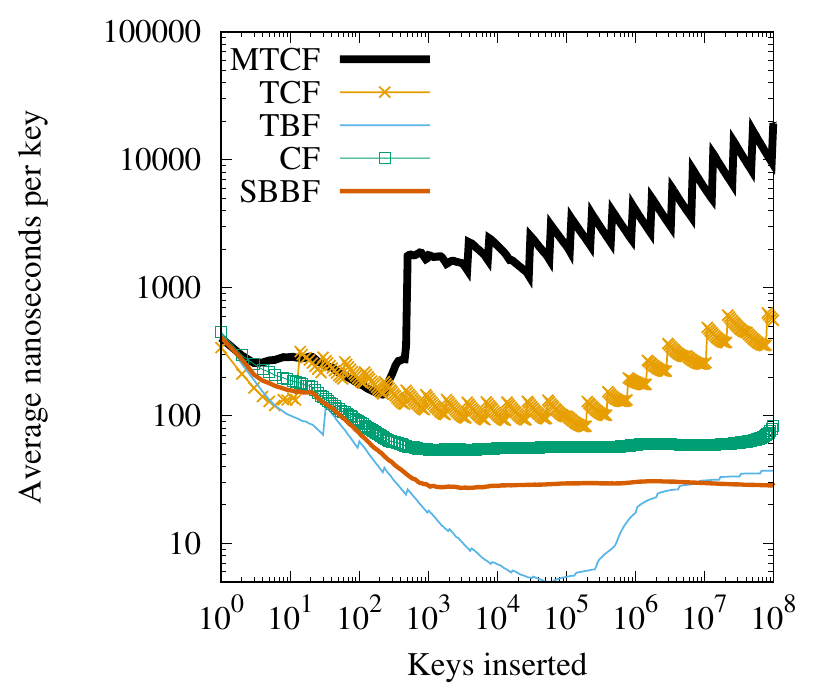}
  \caption{
    \label{insert-time}
    Insert times for filters, i7-7800X.
  }
\end{figure}

\begin{figure}[b!]
  \includegraphics[width=3.333in]{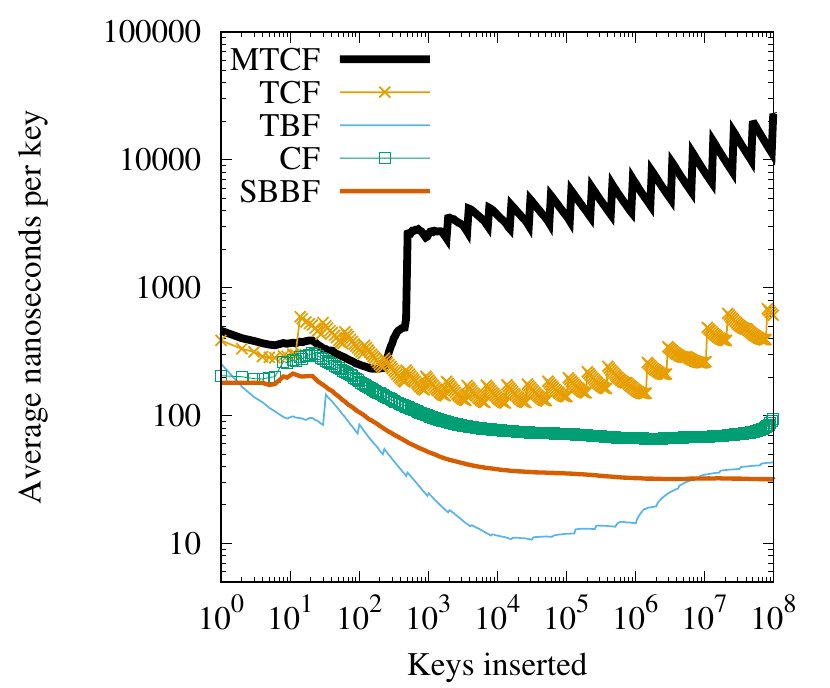}
  \caption{
    \label{arm-insert-time}
    Insert times for filters, m6g.medium.
  }
\end{figure}

\begin{figure}[b!]
  \includegraphics[width=3.333in]{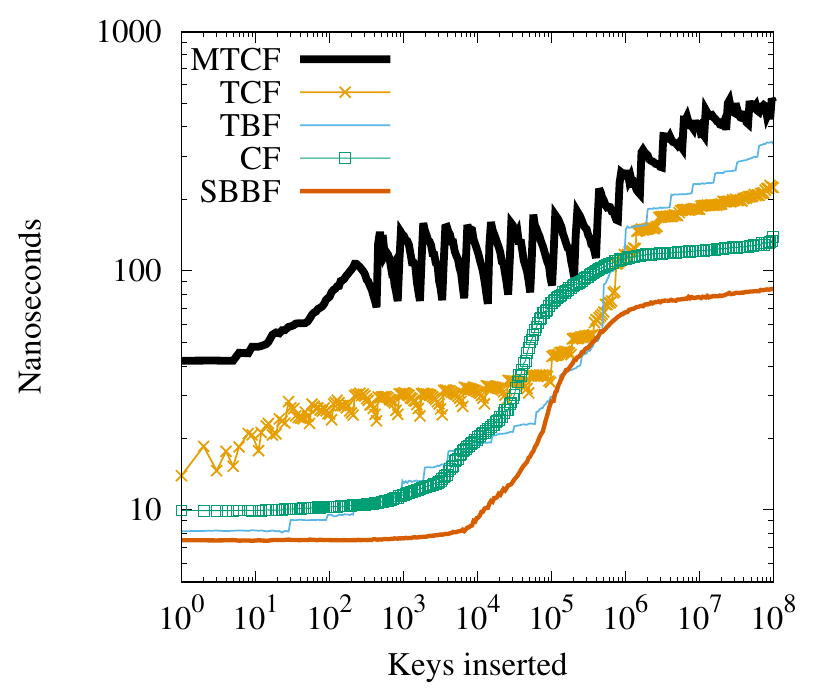}
  \caption{
    \label{lookup-both}
    Lookup times for filters, i7-7800X.
  }
\end{figure}

\begin{figure}[b!]
  \includegraphics[width=3.333in]{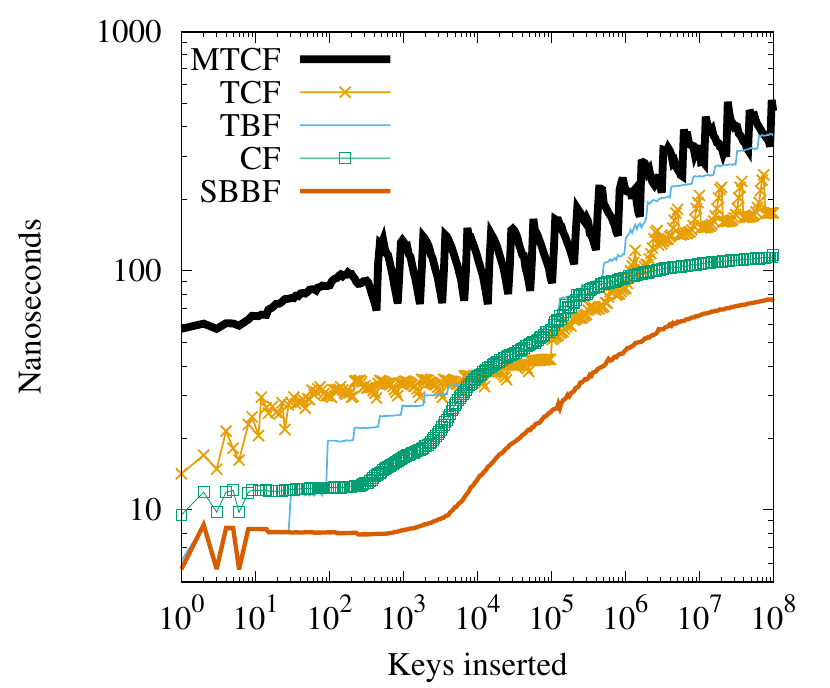}
  \caption{
    \label{arm-lookup-both}
    Lookup times for filters, m6g.medium.
  }
\end{figure}

Figures~\ref{insert-time},~\ref{arm-insert-time},~\ref{lookup-both},~and~\ref{arm-lookup-both} show the performance of \taffy{} filter operations.\footnote{All charts of time show the minimum over nine runs.}
For inserts, \TBF{}s are the fastest of the three \taffy{} filter variants; they are even faster than the fastest non-\taffy{} variant, split block Bloom filters.
\TBF{} inserts are faster than the other cuckoo filters because they are simple, branch-free, and induce a single cache miss; they are faster than the pre-sized split block Bloom filter because, while being built, the entirety of the \TBF{} fits in cache until about 10 million elements have been inserted.
This holds true across both tested machines, x86 and ARM.

For inserts there are visible dips in the average construction time for small filters as they get larger.
These are due to measurement overhead (for the smallest $n$) and the cost of upsizing (for slightly larger $n$).

For lookups, the situation is more complex.
Of the resizable filters, the \taffy{} cuckoo filter is the fastest once the size of the filter is large enough, while a \TBF{} is otherwise faster.
The \MTCF{} lags behind both.

\subsection{Previously-used-password filter}
\label{hibp}

In this section we test the dataset of previously used passwords from ``Have I Been Pwned''~\cite{pwned}.
This dataset consists of 847 million hashes of leaked passwords.
It has grown over time, starting in August 2017 with 306 million passwords. It is currently\footnote{As of December 2021} on version 8.

\Taffy{} Bloom filters and \taffy{} cuckoo filters were tested on this dataset using the 64 low-order bits from the hashes as the keys.
Both filters started out configured with an initial capacity of a single element; the \TBF{} was configured to have a similar fpp to the \TCF{}.
Once insertion was complete, the \TCF{} was frozen to test the lookup performance and fpp of the resulting data structure.
Experiments were conducted on an AWS EC2 r6i.xlarge with 32GiB of memory and an Intel Xeon Platinum 8375C. 
Times are the minimum over a set of 7 runs; fpps are the median.
See Figure~\ref{hibp-table}.

\begin{figure}[b!]
\begin{tabular}{|m{0.5in}|m{0.5in}|m{0.5in}|m{0.5in}|m{0.5in}|}
\hline & {\bf \TBF{}} & {\bf \TCF{}} & {\bf Frozen} & {\bf Raw, sorted}\\
\hline {\bf insert (ns/key)} & 24 & 572 & \TCF{} + 2.2 & 113\\
\hline {\bf fpp} & 0.25\% & 0.26\% & 0.71\% & 0\%\\
\hline {\bf lookup (ns/key)} & 290 & 108 & 70 & 719\\
\hline {\bf space} & 4.1GiB & 4.0GiB & 2.5GiB & 6.3GiB\\
\hline
\end{tabular}
\caption{\label{hibp-table}
Performance on ``Have I Been Pwned''}
\end{figure}

As in the case of the synthetic benchmarks, insert is faster for the \TBF{} and lookup is faster for the \TCF{}.
This omission of the tails makes the false positive rate higher but the lookup faster, since the prefix checks are now unnecessary and the fingerprint matches can now be performed with SIMD-within-a-register techniques, just as in the original cuckoo filter ~\cite{cuckoo-filter-github}.
For the ``Raw, sorted'' column, we consider the cost of storing 64 bits of each hash in a single array with sorting as the input method and binary search as the lookup method.

\subsection{Discussion}

The \MTCF{} offers lower space than the other two \taffy{} filters, but its speed is substantially worse.
It has significant insertion time increases when it is hard to find an eviction sequence; in this case consecutive insert operations may call upsize, causing a spike in the graph. (See Figures~\ref{insert-time}~and~\ref{arm-insert-time}.)
This cyclic behavior was noted by Maier \etal~\cite{dysect}.

During lookup operations on \MTCF{}s, when the cursor is close to 32, the performance improves as the four potential locations to look for a key are more frequently reduced to two, since the shorter permuted keys are no longer long enough for most of the levels in the structure.
See Figures~\ref{lookup-both}~and~\ref{arm-lookup-both}.

Split block Bloom filters and cuckoo filters are still attractive choices when the size of the set to be approximated is known in advance.
When a growable filter is needed, the application matters quite a bit.
If saving every byte matters, \MTCF{}s are called for.
If satellite data (as in a Bloomier filter) is needed, such as when using the filter in front of an LSM tree, a \TCF{} or \MTCF{} should be used, as \TBF{}s do not support satellite data.
Otherwise, a practitioner must ask themselves:

\begin{itemize}
\item Is the workload write-heavy or read-heavy?
  Write-heavy workloads favor \TBF{}s over \TCF{}s.
\item Is the set likely to exceed one million elements (x86) or 1000 elements (ARM)?
  If yes, a \TCF{} should be preferred.
\end{itemize}

\ifanon
\else
The code for \taffy{} filters is available \ifanon\else on GitHub\fi under a permissive open-source license.\ifanon\else\footnote{\url{https://github.com/jbapple/libfilter}}\fi
\fi








\section{Conclusion}
\label{conclusion}

This work exhibits for the first time practical structures supporting approximate membership queries and filter growth without exceeding $O(\lg (1/\varepsilon) + \lg \lg N)$ bits of space used per distinct key.
We presented three structures: the \TBF{}, the \TCF{}, and the \MTCF{}.
We demonstrated taffy filter performance and correctness under synthetic and real-world benchmarks. 







\ifanon
\else
\section*{Acknowledgments}
Thanks to Pedro Vasallo and Alex Breslow for helpful discussions and feedback.
\fi

\bibliographystyle{plain}
\bibliography{taffy}

\end{document}